\def\be{\begin{equation}}
\def\ee{\end{equation}}
\def\ba{\begin{array}}
\def\ea{\end{array}}

\documentclass[prl,showpacs,twocolumn,amsmath]{revtex4}
\usepackage{mathrsfs}
\usepackage{amssymb}
\usepackage{pifont}
\usepackage{CJKutf8}
\usepackage{float}  
\usepackage{epsfig,subfigure,dsfont,amsthm,amsbsy,mathrsfs,amscd}
\def\qed{\leavevmode\unskip\penalty9999 \hbox{}\nobreak\hfill
     \quad\hbox{\leavevmode  \hbox to.77778em{%
               \hfil\vrule   \vbox to.675em%
               {\hrule width.6em\vfil\hrule}\vrule\hfil}}
     \par\vskip3pt}

\newtheorem{theorem}{Theorem}

\newtheorem{lemma}{Lemma}
\input amssym.def
\newenvironment{shrinkeq}[1]
{ \bgroup
  \addtolength\abovedisplayshortskip{#1}
  \addtolength\abovedisplayskip{#1}
  \addtolength\belowdisplayshortskip{#1}
  \addtolength\belowdisplayskip{#1}}
{\egroup\ignorespacesafterend}
\begin{document}
\title{\large\bf Notes on quantum coherence with $l_1$-norm and convex-roof $l_1$-norm}
\author{Jiayao Zhu$^{1}$, Jian Ma$^{1, 2}$, Tinggui Zhang$^{1, 2, \dag}$}
\affiliation{ $^{1}$School of Mathematics and Statistics, Hainan
Normal University, Haikou, 571158, China\\
$^{2}$ Key Laboratory of Data Science and Smart Education, Ministry of Education, Hainan Normal University, Haikou 571158, China \\
 \footnotesize\small $^{\dag}$ Correspondence to
tinggui333@163.com}
\date{17 July, 2021}
\bigskip

\begin{abstract}
{\bf Abstract} In this work, we evaluate quantum coherence using the
$l_{1}$-norm and convex-roof $l_{1}$-norm and obtain several new
results. First, we provide some new general triangle-like inequalities
of quantum coherence, with results better than existing
ones. Second, for some special three-dimensional quantum states, a
method for calculating the convex-roof $l_{1}$-norm is presented. Lastly, we offer distinct upper bounds in the $l_{1}$-norm measure of
coherence based on the quantum state itself.

{\bf Keywords} Quantum coherence; $l_{1}$-norm; Triangle-like
inequalities; Convex-roof $l_{1}$-norm
\end{abstract}

\pacs{03.67.-a, 02.20.Hj, 03.65.-w}
 \maketitle
\bigskip
\section{Introduction }
Quantum coherence stems from the superposition of quantum states; it reflects the ability of quantum states to display quantum coherence effects.
It is the most basic property distinguishing quantum physics from classical physics, and it is also an important quantum mechanical property that is widely used in quantum
 information processing and quantum computing. Consequently, quantum information science and the derived quantum technology offer advantages that the corresponding
 classical informatics and technology cannot provide, such as the theoretical security of quantum communication\cite{1}, ultra-fast quantum computing\cite{2}, and quantum precision measurements with accuracies exceeding the classical limit\cite{3}. Additionally,
 quantum coherence is widely used in quantum biology\cite{4,5}, quantum thermodynamics\cite{6,7}, and quantum computing\cite{8,9}.
 Therefore, quantifying the quantum coherence in a resource framework is important and meaningful, and the study of quantum coherence plays an important role in furthering the frontiers of physics research.

Quantum coherence and entanglement are two characteristics of the quantum world. Quantum entanglement describes bipartite or multiple systems, whereas quantum coherence is defined for single systems.
With the rapid developments in quantum information science, researchers have found that, similar to entanglement, quantum coherence can be treated as a physical resource. However, a core issue in this field is the
quantification of coherence. Adopting some of the methods used with entanglement, Baumgratz et al. established a
quantitative theory of coherence as a resource along with four necessary conditions that should be satisfied by any appropriate measure of coherence \cite{10}.

Most of the measures of quantum coherence are based on distances, defined in terms of the minimum distance between the selected quantum state and the given
set of incoherent states $\tilde{\mathcal{I}}$. These incoherent states are quantum states that are diagonalizable under a certain reference basis $\left\{ {|i\rangle } \right\}_{i = 1}^d$ in
the Hilbert space, that is, $\delta  = \sum\nolimits_{i = 1}^d {{\delta _i}|i\rangle \langle i|}$. A measure of coherence is defined as follows:
$${C_D}\left( \rho  \right) = \mathop {\min }\limits_{\delta  \in
\tilde{\mathcal{I}}} D\left( {\rho ,\delta } \right),$$ where
$D\left( {\rho ,\delta } \right)$ denotes a certain measure of the
distance between $\rho$ and $\delta$.

Baumgratz and his coworkers found that the relative entropy of coherence
$$C_{re}(\rho ) = S(\rho _{diag}) - S(\rho ),$$
where $S$ is the von Neumann entropy and ${\rho _{diag}}$ denotes the state obtained from $\rho$ by deleting all off-diagonal elements and the $l_{1}$-norm of coherence
\begin{equation*}
     {C_{{l_1}}}(\rho ) = \sum\limits_{i \ne j} {\left| {{\rho _{ij}}} \right|},
\end{equation*}
are both appropriate measures of coherence.

Thus far, many suitable measures have been reported, such as the
robustness of coherence\cite{11}, maximum relative entropy of
coherence\cite{12}, entanglement-based coherence
measurement\cite{13}, average quantum coherence\cite{mjmt}, and set
coherence \cite{sdru}. However, the $l_{1}$-norm of coherence has
continued to be significant for research. Recently, it was
discovered that the $l_{1}$-norm of coherence can be used to
describe wave-particle duality \cite{fcle}.

Ref.\cite{14} showed that, if a rank-2 state $\rho$ can be expressed
as a convex combination of two pure states, i.e.
\begin{equation*}
       \rho = {p_1}|{\psi _1}\rangle \langle {\psi _1}| + {p_2}|{\psi _2}\rangle \langle {\psi
       _2}|,
\end{equation*}
a triangle-like inequality can be established as follows:
\begin{equation*}
\begin{aligned}
& \left| {E\left( {\sqrt {{p_1}} |{\psi _1}\rangle } \right) - E\left( {\sqrt {{p_2}} |{\psi _2}\rangle } \right)} \right|  \le E(\rho ) \\
&\qquad \qquad \qquad \qquad \le E\left( {\sqrt {{p_1}} |{\psi
_1}\rangle } \right) + E\left( {\sqrt {{p_2}} |{\psi _2}\rangle }
\right),
\end{aligned}
\end{equation*}
where $E$ can be considered as either a coherence measure or an entanglement concurrence.

Ref.\cite{15} provided a general triangle-like inequality satisfied
by the $l_{1}$-norm measure of coherence for a convex combination of
$n$ arbitrary pure states of a quantum state $\rho$, thus verifying
the conclusions presented in Ref.\cite{14}.

Both Ref.\cite{14} and Ref.\cite{15} estimated the coherence of a
quantum state by using state decomposition. We can also evaluate the
coherence of a quantum state based on its properties. As mentioned
in Ref.\cite{16}, for a $d$-dimensional quantum state, its
$l_{1}$-norm is not more than $d - 1$.

To date, many scholars have estimated the coherence of a quantum state, providing lower and upper bounds in different forms. However, we find that some of
these estimates can be optimized, and we can measure the coherence of a quantum state from different perspectives, thus affording distinct conclusions. In this study, we select the $l_{1}$-norm and convex-roof $l_{1}$-norm
to quantify coherence. Based on the state decomposition of a quantum state and its properties, we provide some new lower and upper bounds for  the $l_{1}$-norm and convex-roof $l_{1}$-norm.

\section{ Generalized triangular inequality for quantum coherence based on $l_1$ norm }
On basis of the state decomposition of a quantum state,
Ref.\cite{14} provides the following conclusion:

\begin{lemma} \label{l2}
If a quantum state $\rho$ can be expressed as a convex combination
of two states, that is, $\rho = p_{1}\rho_{1} + p_{2}\rho_{2}$, then we
have
\begin{equation*}
        \left| {{p_1}{C_{{l_1}}}({\rho _1}) - {p_2}{C_{{l_1}}}({\rho _2})} \right| \le {C_{{l_1}}}(\rho ).
\end{equation*}
\end{lemma}

Moreover, Ref.\cite{15} provides a triangle-like inequality in the
$l_{1}$-norm measure of coherence.
\begin{lemma} \label{l1}
If the state $\rho$ can be expressed as a convex combination of $n(n \geq 2)$ states, that is, $\rho  = \sum\nolimits_{i = 1}^n {{p_i}{\rho _i}}$, then ${C_{{l_1}}}\left( \rho  \right)$ satisfies the
 following triangle-like inequality
\begin{equation*}
   \frac{1}{n}\sum\limits_{k = 1}^n {\left| {G_k^{(n - 1)} - {p_k}{C_{{l_1}}}({\rho _k})} \right|}  \le {C_{{l_1}}}\left( \rho  \right) \le \sum\limits_{k = 1}^n {{p_k}{C_{{l_1}}}({\rho
   _k})},
\end{equation*}
where $G_k^{(n - 1)},k = 1,2, \cdots ,n$ are the lower bounds of $\sum\nolimits_{i \ne k}^n {{p_i}{C_{{l_1}}}} \left( {\frac{{\sum\nolimits_{j \ne k}^n {{p_j}{\rho _j}} }}{{\sum\nolimits_{t \ne k}^n {{p_t}} }}} \right)$.
\end{lemma}

We can further develop the conclusion of Lemma 2. However, we first introduce a useful lemma followed by a new
triangle-like inequality regarding the $l_{1}$-norm considering the
state decomposition of a quantum state.

\begin{lemma}  \label{l4}
Let ${a_i} \ge 0(i = 1,2, \cdots ,n)$ and not be complete zeros. If $b \ge {a_i},i = 1,2, \cdots ,n$, then we obtain 
\begin{equation*}
\begin{aligned}
&b \ge \frac{{\sum\nolimits_{i = 1}^n {a_i^2} }}{{\sum\nolimits_{i = 1}^n {a_i^{}} }} \ge \frac{1}{n}\sum\nolimits_{i = 1}^n {a_i^{}}.
\end{aligned}
\end{equation*}
\end{lemma}

{\bf{Proof:}} Based on these assumptions, we have
\begin{equation*}
\begin{aligned}
&\sum\nolimits_{i = 1}^n {a_i^2}  \le \sum\nolimits_{i = 1}^n {a_i^{}} b = b\sum\nolimits_{i = 1}^n {a_i^{}}.
\end{aligned}
\end{equation*}

Note that ${a_1}, \cdots ,{a_n}$ are not complete zeros; thus,
$\sum\nolimits_{i = 1}^n {a_i} > 0$. Therefore, it holds that
\begin{equation*}
\begin{aligned}
&b \ge \frac{{\sum\nolimits_{i = 1}^n {a_i^2} }}{{\sum\nolimits_{i = 1}^n {a_i^{}} }}.
\end{aligned}
\end{equation*}
Moreover, using the inequality
\begin{equation*}
\begin{aligned}
&{({x_1} + {x_2} +  \cdots  + {x_n})^2} \le n(x_{\rm{1}}^{\rm{2}} + x_{\rm{2}}^{\rm{2}} +  \cdots  + x_n^2),
\end{aligned}
\end{equation*}
we can directly obtain
\begin{equation*}
\begin{aligned}
&\frac{{\sum\nolimits_{i = 1}^n {a_i^2} }}{{\sum\nolimits_{i = 1}^n {a_i^{}} }} \ge \frac{1}{n}\sum\nolimits_{i = 1}^n {a_i^{}} .
\end{aligned}
\end{equation*}
\hfill \rule{1ex}{1ex}

Next, we present the general triangle-like inequality based on the $l_{1}$-norm measure of coherence.
\begin{theorem}
If the state $\rho$ can be expressed as a convex combination of $n(n \geq 2)$ states, that is, $\rho  = \sum\nolimits_{i = 1}^n {{p_i}{\rho _i}}$, then ${C_{{l_1}}}(\rho )$ satisfies
\begin{equation}
\frac{{\sum\nolimits_{i = 1}^n {A_i^2} }}{{\sum\nolimits_{i = 1}^n
{{A_i}} }} \le {C_{{l_1}}}(\rho ) \le
\frac{{\rm{1}}}{n}\sum\limits_{k = 1}^n {\left[ {G_k^{(n - 1)} +
{p_k}{C_{{l_1}}}({\rho _k})} \right]}, \label{1}
\end{equation}
where
\begin{equation*}
  G_k^{(n - 1)} = (1 - {p_k}){C_{{l_1}}}({\textstyle{\frac{\rho  - {p_k}{\rho _k}}{1 - {p_k}}}}),k = 1,2, \cdots,
  ,n;
\end{equation*}
\begin{equation*}
   {A_i} = \left| {G_i^{(n - 1)} - {p_i}{C_{{l_1}}}({\rho _i})} \right|.
\end{equation*}
\end{theorem}

It should be noted that we only consider the case where
\begin{equation*}
  G_k^{(n - 1)} - {p_k}{C_{{l_1}}}({\rho _k}), k = 1,2, \cdots ,n
\end{equation*}
are not all zeros. If
\begin{equation*}
  G_k^{(n - 1)} - {p_k}{C_{{l_1}}}({\rho _k}) = 0, k = 1,2, \cdots ,n,
\end{equation*}
then 0 can act as a lower bound of ${C_{{l_1}}}(\rho )$.

{\bf{Proof:}} First, we analyze the right-hand side of Eq.(\ref{1}). In
the case of $n = 2$, it holds that
\begin{equation}
  G_1^{(1)} = {p_2}{C_{{l_1}}}({\rho _2}),  G_2^{(1)} = {p_1}{C_{{l_1}}}({\rho _1}). \label{2}
\end{equation}
Therefore, from Lemma 2, we have
\begin{equation*}
\begin{aligned}
&\quad \  \frac{1}{2}\left[ {(G_1^{(1)} + {p_1}{C_{{l_1}}}({\rho _1})) + (G_2^{(1)} + {p_2}{C_{{l_1}}}({\rho _2}))} \right] \\
&= \frac{1}{2}\left[ {2({p_1}{C_{{l_1}}}({\rho _1}) + {p_2}{C_{{l_1}}}({\rho _2}))} \right]\\
& = {p_1}{C_{{l_1}}}({\rho _1}) + {p_2}{C_{{l_1}}}({\rho _2}) \ge {C_{{l_1}}}(\rho ).
\end{aligned}
\end{equation*}
Hence, the conclusion is true when $n = 2$.

Assuming that the conclusion is true when $n = m$,
let us analyze the case of $n = m + 1$. For $\forall k \in \{ 1,2,
\cdots ,m + 1\}$,
\begin{equation*}
\begin{aligned}
&\sum\limits_{i \ne k} {{\textstyle{\frac{{p_i}}{1 - {p_k}}}}{\rho _i}}
\end{aligned}
\end{equation*}
remains a density matrix. According to Lemma 2, we have
\begin{equation*}
\begin{aligned}
{C_{{l_1}}}(\rho )& = {C_{{l_1}}}((1 - {p_k})(\sum\limits_{i \ne k} {{\textstyle{\frac{{p_i}}{1 - {p_k}}}}{\rho _i}}) + {p_k}{\rho _k}) \\
&\le (1 - {p_k}){C_{{l_1}}}(\sum\limits_{i \ne k} {{\textstyle{{\frac{p_i}{1 - {p_k}}}}{\rho _i}}} ) + {p_k}{C_{{l_1}}}({\rho _k}) \\
&= G_k^{(m)} + {p_k}{C_{{l_1}}}({\rho _k}).
\end{aligned}
\end{equation*}
For any arbitrary $k$, it holds that
\begin{equation*}
\begin{aligned}
&(m + 1){C_{{l_1}}}(\rho ) = \sum\limits_{k = 1}^{m + 1}
{{C_{{l_1}}}(\rho )}  \le \sum\limits_{k = 1}^{m + 1} (G_k^{(m)} +
{p_k}{C_{{l_1}}}({\rho _k})).
\end{aligned}
\end{equation*}
In other words,
\begin{equation*}
\begin{aligned}
&{C_{{l_1}}}(\rho ) \le \frac{1}{{m + 1}}\sum\limits_{k = 1}^{m + 1}
(G_k^{(m)} + {p_k}{C_{{l_1}}}({\rho _k})).
\end{aligned}
\end{equation*}
Hence, the inequality also holds when $n = m + 1$. Thus
the right side of Eq.(\ref{1}) is true.

Next, we consider the left side of Eq.(\ref{1}).

First, we analyze the case of $n = 2$. From Eq.$(\ref{2})$, we have
\begin{equation*}
\begin{aligned}
&\frac{{{{\left| {{p_2}{C_{{l_1}}}({\rho _2}) - {p_1}{C_{{l_1}}}({\rho _1})} \right|}^2} + {{\left| {{p_1}{C_{{l_1}}}({\rho _1}) -
 {p_2}{C_{{l_1}}}({\rho _2})} \right|}^2}}}{{2\left| {{p_1}{C_{{l_1}}}({\rho _1}) - {p_2}{C_{{l_1}}}({\rho _2})} \right|}}
\end{aligned}
\end{equation*}
\begin{equation*}
\begin{aligned}
& =\left| {{p_1}{C_{{l_1}}}({\rho _1}) - {p_2}{C_{{l_1}}}({\rho
_2})} \right|
  \le {C_{{l_1}}}(\rho ).
\end{aligned}
\end{equation*}
Thus, the conclusion holds when $n = 2$.

Assuming that the conclusion holds when $n = m$, we now consider the case of $n = m + 1$.

For $\forall k \in \{ 1,2, \cdots ,m + 1\}$, according to Lemma 1,
we have
\begin{equation*}
\begin{aligned}
{C_{{l_1}}}(\rho ) &= {C_{{l_1}}}((1 - {p_k})(\sum\limits_{i \ne k} {{\textstyle{\frac{{p_i}} {1 - {p_k}}}}{\rho _i}})  + {p_k}{\rho _k}) \\
&\ge \left| {(1 - {p_k}){C_{{l_1}}}(\sum\limits_{i \ne k}
{{\textstyle{\frac{{p_i}}{1 - {p_k}}}}{\rho _i}} ) -
{p_k}{C_{{l_1}}}
({\rho _k})} \right|  \\
& = \left| {G_k^{(m)} - {p_k}{C_{{l_1}}}({\rho _k})} \right|.
\end{aligned}
\end{equation*}
Based on Lemma 3, the conclusion is true for the
case of $n = m + 1$.

Combined with the above analysis, Eq.(\ref{1}) is true for any
arbitrary natural number $n$. \hfill \rule{1ex}{1ex}

Furthermore, we can prove that our estimate of coherence
is more accurate than that provided by Lemma 2.

\begin{theorem}
If $\rho$ can be expressed as a convex combination of $n$ states, that is, $\rho  = \sum\nolimits_{i = 1}^n {{p_i}{\rho _i}}$, then
\begin{shrinkeq}{-1.8ex}
\begin{equation}
\frac{1}{n}\sum\limits_{k = 1}^n {\left| {G_k^{(n - 1)} - {p_k}{C_{{l_1}}}({\rho _k})} \right|}  \le \frac{{\sum\nolimits_{i = 1}^n {A_i^2} }}{{\sum\nolimits_{i = 1}^n {{A_i}} }},  \label{3}
\end{equation}
\end{shrinkeq}
\begin{shrinkeq}{-1.8ex}
\begin{equation}
\frac{1}{n}\sum\limits_{k = 1}^n (G_k^{(n - 1)} +
{p_k}{C_{{l_1}}}({\rho _k}))  \le \sum\limits_{k = 1}^n
{{p_k}{C_{{l_1}}} ({\rho _k})}, \label{4}
\end{equation}
\end{shrinkeq}
where ${A_i} = |G_i^{(n - 1)} - {p_i}{C_{{l_1}}}({\rho _i})|,i = 1,2, \cdots ,n$.
\end{theorem}

{\bf{Proof:}} Assume that $\rho$ can be expressed as a convex
combination of $n$ states, i.e., $\rho  = \sum\nolimits_{i = 1}^n
{{p_i}{\rho _i}}$. Based on Lemma 3, Eq.(\ref{3}) is
true.

We also have
\begin{equation*}
\begin{aligned}
&\quad \ \frac{1}{n}\sum\limits_{k = 1}^n (G_k^{(n - 1)} + {p_k}{C_{{l_1}}}({\rho _k})) \\
& = \frac{1}{n}\sum\limits_{k = 1}^n {G_k^{(n - 1)}}  + \frac{1}{n}\sum\limits_{k = 1}^n {{p_k}{C_{{l_1}}}({\rho _k})} \\
& = \frac{1}{n}\sum\limits_{k = 1}^n {(1 - {p_k}){C_{{l_1}}}(\sum\limits_{i \ne k} {{\textstyle{\frac{{p_i}}{1 - {p_k}}}}{\rho _i}} )}  + \frac{1}{n}\sum\limits_{k = 1}^n {{p_k}{C_{{l_1}}}({\rho _k})}  \\
&  \le \frac{1}{n}\sum\limits_{k = 1}^n {(1 - {p_k})\sum\limits_{i \ne k} {{\textstyle{\frac{{p_i}}{1 - {p_k}}}}} {C_{{l_1}}}({\rho _i})}  + \frac{1}{n}\sum\limits_{k = 1}^n {{p_k}{C_{{l_1}}}({\rho _k})} \\
& = \frac{1}{n}\sum\limits_{k = 1}^n {\sum\limits_{i \ne k} p_i {{C_{{l_1}}}({\rho _i})} }  + \frac{1}{n}\sum\limits_{k = 1}^n {{p_k}{C_{{l_1}}}({\rho _k})}  \\
& = \frac{(n - 1)}{n}\sum\limits_{k = 1}^n {{p_k}{C_{{l_1}}}({\rho _k})}  + \frac{1}{n}\sum\limits_{k = 1}^n {{p_k}{C_{{l_1}}}({\rho _k})} \\
& = \sum\limits_{k = 1}^n {{p_k}{C_{{l_1}}}({\rho _k})}.
\end{aligned}
\end{equation*}
Thus, Eq.(\ref{4}) holds. \hfill \rule{1ex}{1ex}

\section{ Generalized triangular inequality for quantum coherence based on convex-roof $l_{1}$-norm }

The convex-roof $l_{1}$-norm is another method for measuring
coherence\cite{17}. The convex-roof $l_{1}$-norm of a mixed state is
defined as
\begin{equation*}
\begin{aligned}
&\widetilde {{C_{{l_1}}}}(\rho ) = \mathop {\min }\limits_{\{ {p_i},|{\psi _i}\rangle \} } \sum\limits_i {{p_i}{C_{{l_1}}}(|{\psi _i}\rangle )}.
\end{aligned}
\end{equation*}

However, we are unfamiliar with the convex-roof $l_{1}$-norm owing
to the difficulty in its calculation. To date, we can only calculate
the convex-roof $l_{1}$-norm of a two-dimensional state\cite{17,18}
and some special high-dimensional states\cite{19}. In this study, we
prove that, for certain three-dimensional quantum states,
${C_{{l_1}}}\left( \rho \right) = \widetilde {{C_{{l_1}}}}\left(
\rho  \right)$ holds. We also provide a distinct triangle-like
inequality in the convex-roof $l_{1}$-norm of coherence. To this
end, we begin with the two lemmas mentioned in Ref.\cite{17} and
Ref.\cite{14}.
\begin{lemma} \label{l5}
For any quantum state $\rho$, we have
\begin{equation*}
     {C_{{l_1}}}(\rho ) \le \widetilde {{C_{{l_1}}}}(\rho ).
\end{equation*}
\end{lemma}

\begin{lemma} \label{l6}
If $\rho$ is a pure state, then
\begin{equation*}
     {C_{{l_1}}}(\rho ) = \widetilde {{C_{{l_1}}}}(\rho ).
\end{equation*}
\end{lemma}

According to Lemma \ref{l5}, we obtain the following theorem.
\begin{theorem}  \label{dl3} 
Let $\rho$ be a three-dimensional quantum state with $rank(\rho ) = 3$.
If the matrix of $\rho$ under reference basis $\left\{ {|i\rangle }
\right\}_{i = 1}^d$ has the following form
\begin{equation*}
\begin{aligned}
&\rho  = \left( {\begin{array}{*{20}{c}}
{{\rho _{11}}}&{{\rho _{12}}}&0\\
{\overline {{\rho _{12}}} }&{{\rho _{22}}}&0\\
0&0&{{\rho _{33}}}
\end{array}} \right),
\end{aligned}
\end{equation*}
then
\begin{equation*}
  {C_{{l_1}}}(\rho ) = \widetilde {{C_{{l_1}}}}(\rho ).
\end{equation*}
\end{theorem}

{\bf{Proof:}}
Select $|{\psi _1}\rangle ,|{\psi _2}\rangle ,|{\psi _3}\rangle$, and ${p_1},{p_2},{p_3}$ as follows:
\begin{shrinkeq}{-1ex}
\begin{equation*}
|{\psi _1}\rangle  = \frac{1}{{\sqrt {{p_1}} }}\left( {\sqrt {{\rho _{11}}} |1\rangle  + \frac{{\left| {{\rho _{12}}} \right|}}{{\sqrt {{\rho _{11}}} }}{e^{ - iarg{\rho _{12}}}}|2\rangle } \right),
\end{equation*}
\begin{equation*}
      |{\psi _2}\rangle  = |2\rangle ,|{\psi _3}\rangle  = |3\rangle,  \\
\end{equation*}
\begin{equation*}
{p_1} = {\rho _{11}} + \frac{{{{\left| {{\rho _{12}}} \right|}^2}}}{{{\rho _{11}}}},{p_2} = {\rho _{22}} - \frac{{{{\left| {{\rho _{12}}} \right|}^2}}}{{{\rho _{11}}}},{p_3} = {\rho _{33}}.
\end{equation*}
\end{shrinkeq}
Thus,
\begin{equation*}
  \rho  = {p_1}|{\psi _1}\rangle \langle {\psi _1}| + {p_2}|{\psi _2}\rangle \langle {\psi _2}| + {p_3}|{\psi _3}\rangle \langle {\psi _3}|.
\end{equation*}
and
\begin{equation*}
 {p_1}{C_{{l_1}}}\left( {|{\psi _1}\rangle } \right) + {p_2}{C_{{l_1}}}\left( {|{\psi _2}\rangle } \right) + {p_3}{C_{{l_1}}}\left( {|{\psi _3}\rangle } \right) = {C_{{l_1}}}\left( \rho  \right).
\end{equation*}
According to the definition of the convex-roof $l_{1}$-norm, we have
\begin{equation*}
  {C_{{l_1}}}(\rho ) \geq \widetilde {{C_{{l_1}}}}(\rho ).
\end{equation*}
Thus, it holds that
\begin{equation*}
  {C_{{l_1}}}(\rho ) = \widetilde {{C_{{l_1}}}}(\rho ).
\end{equation*}
\hfill \rule{1ex}{1ex}

Ref.\cite{15} provides the following triangle-like inequality regarding the
convex-roof $l_{1}$-norm by using pure state decomposition.
\begin{lemma} \label{yl8} 
If $\rho$ can be expressed as a convex combination of $n$ linearly independent pure states, that is, $\rho  = \sum\nolimits_{i = 1}^n {{p_i}|{\psi _i}\rangle \langle {\psi _i}|} $, then the convex-roof $l_{1}$-norm
satisfies the following triangle-like inequality
\begin{equation}
  \frac{1}{n}\sum\limits_{k = 1}^n {\left| {\widetilde G_k^{(n - 1)} - {p_k}\widetilde {{C_{{l_1}}}}\left( {|{\psi _k}\rangle } \right)} \right| \le } \widetilde {{C_{{l_1}}}}\left( \rho  \right) \le \sum\limits_i {{p_i}\widetilde {{C_{{l_1}}}}\left( {|{\psi _i}\rangle } \right)}, \label{jzx}
\end{equation}
where $\widetilde {{C_{{l_1}}}}\left( {|{\psi _i}\rangle } \right) = {C_{{l_1}}}\left( {|{\psi _i}\rangle } \right)$ is the convex-roof $l_{1}$-norm of the pure state $|{\psi _i}\rangle \left( {1 \le i \le n} \right)$ and
 $\widetilde G_k^{(n - 1)}\left( {1 \le k \le n} \right)$ is the lower bound of $\sum\nolimits_{i \ne k}^n {{p_i}\widetilde {{C_{{l_1}}}}} \left( {\frac{{\sum\nolimits_{j \ne k}^n {{p_j}{\rho _j}} }}{{\sum\nolimits_{t \ne k}^n {{p_t}} }}} \right)$.
\end{lemma}

More importantly, while the correctness of the left side of
Eq.(\ref{jzx}) can be discussed, the right side of Eq.(\ref{jzx})
can be improved.

Next, we present a new general triangle-like inequality for the convex-roof $l_{1}$-norm measure of coherence.
\begin{theorem} \label{dl4}  
Let $\rho$ be a mixed state that can be expressed as a convex
combination of $n\left( {n \ge 2} \right)$ linearly independent pure
states: $|{\psi _1}\rangle ,|{\psi _2}\rangle , \cdots ,|{\psi
_n}\rangle$, i.e. $\rho  = \sum\nolimits_{i = 1}^n {{p_i}|{\psi
_i}\rangle \langle {\psi _i}|}$. If we know some off-diagonal
element $\rho {}_{st}$ of $\rho$, then
\begin{equation}
\frac{{{{\left| {{\rho _{st}}} \right|}^2}}}{B} \le \widetilde
{{C_{{l_1}}}}(\rho ) \le \frac{1}{n}\sum\limits_{k = 1}^n {\left[
{\widetilde G_k^{(n - 1)} + {p_k}\widetilde {{C_{{l_1}}}}(|{\psi
_k}\rangle )} \right]},  \label{6}
\end{equation}
where
\begin{equation*}
  B = {{n(\sum\limits_{k = 1}^n {{p_k}{C_{{l_1}}}(|{\psi _k}\rangle )}  + 1)}},
\end{equation*}
\begin{equation*}
  \widetilde G_k^{(n - 1)} = (1 - {p_k})\widetilde {{C_{{l_1}}}}(\sum\limits_{i \ne k} {{\textstyle{\frac{{p_i}}{1 - {p_k}}}}|{\psi _k}\rangle )} ,k = 1,\cdots ,n.
\end{equation*}
\end{theorem}

{\bf{Proof:}}
We use mathematical induction to prove this theorem. First, we consider the case of $n = 2$. Assume that $\rho  = {p_1}{\rho _1} + {p_2}{\rho _2}$, where ${\rho _1} and {\rho _2}$ are
 two linearly independent pure states. According to the definition of the convex-roof $l_{1}$-norm, we have
\begin{equation*}
\begin{aligned}
&{\widetilde C_{{l_1}}}(\rho ) \le {p_1}\widetilde {{C_{{l_1}}}}({\rho _1}) + {p_2}\widetilde {{C_{{l_1}}}}({\rho _2}).
\end{aligned}
\end{equation*}

According to the definition, we can directly obtain
\begin{equation*}
  \widetilde G_1^{(1)} = {p_2}\widetilde {{C_{{l_1}}}}({\rho _2}),\widetilde G_2^{(1)} = {p_1}\widetilde {{C_{{l_1}}}}({\rho _1}).
\end{equation*}
Furthermore, we have
\begin{equation*}
\begin{aligned}
{\widetilde C_{{l_1}}}(\rho ) &\le {p_1}\widetilde {{C_{{l_1}}}}({\rho _1}) + {p_2}\widetilde {{C_{{l_1}}}}({\rho _2}) \\
& =\frac{1} {2}\left[ {(\widetilde G_1^{(1)} + {p_1}\widetilde
{{C_{{l_1}}}}({\rho _1})) + (\widetilde G_2^{(1)} +
 {p_2}\widetilde {{C_{{l_1}}}}({\rho _2}))} \right].
\end{aligned}
\end{equation*}
Therefore, the conclusion is true when $n = 2$.

Assuming that the conclusion holds in the case of $n = m$, we now analyze the case of $n = m + 1$.

For $\forall k \in \{ 1,2, \cdots ,m + 1\}$, we have
\begin{equation}
\begin{aligned}
\widetilde {{C_{{l_1}}}}(\rho )& \le (1 - {p_k})\widetilde {{C_{{l_1}}}}(\sum\limits_{i \ne k} {{\textstyle{\frac{{p_i}}
 {1 - {p_k}}}}{\rho _i}} ) + {p_k}\widetilde {{C_{{l_1}}}}({\rho _k}) \\
& = \widetilde G_k^{(n - 1)} + {p_k} \widetilde {{C_{{l_1}}}}({\rho _k}).  \label{7}
\end{aligned}
\end{equation}
Summing over all the $k$ in Eq.(\ref{7}), we have
\begin{equation*}
\begin{aligned}
&\widetilde {{C_{{l_1}}}}(\rho ) \le \frac{1}{{m + 1}}\sum\limits_{k
= 1}^{m + 1} (\widetilde G_k^{(m)} + {p_k}\widetilde
{{C_{{l_1}}}}({\rho _k})).
\end{aligned}
\end{equation*}
In other words, the conclusion is true in the case of $n = m + 1$. Thus, the
right side of Eq.(\ref{6}) holds.

Next, we analyze the left side of Eq.
(\ref{6}). Assume that $\left\{
{{q_k},|{\phi _k}\rangle } \right\}$ is the optimal decomposition of
$\rho$, i.e.,
\begin{equation*}
\begin{aligned}
&\widetilde {{C_{{l_1}}}}(\rho ) = \sum\limits_{k = 1}^n {{q_k}{C_{{l_1}}}(|{\phi _k}\rangle )}.
\end{aligned}
\end{equation*}
Let $|{\phi _k}\rangle  = \sum\nolimits_{j = 1}^d {{\phi _{kj}}{e^{i{\theta _{kj}}}}|j\rangle } ,k = 1,2, \cdots ,n$, where ${\phi _{kj}}$ are non-negative real numbers and ${\theta _{kj}}$ are
real numbers. Thus,
\begin{equation*}
\begin{aligned}
|{\phi _k}\rangle \langle {\phi _k}|& = \left( {\sum\nolimits_{j = 1}^d {{\phi _{kj}}{e^{i{\theta _{kj}}}}|j\rangle } } \right)\left( {\sum\nolimits_{t = 1}^d {{\phi _{kt}}{e^{ - i{\theta _{kt}}}}\langle t|} } \right) \\
&  = \left( {\sum\nolimits_{j = 1}^d {\sum\nolimits_{t = 1}^d {{\phi _{kj}}{\phi _{kt}}{e^{i({\theta _{kj}} - {\theta _{kt}})}}|j\rangle \langle t|} } } \right).
\end{aligned}
\end{equation*}
Therefore,
\begin{shrinkeq}{-1.2ex}
\begin{equation}
  {C_{{l_1}}}(|{\phi _k}\rangle ) = \sum\limits_{i \ne j} {{\phi _{ki}}{\phi _{kj}}}.  \label{8}
\end{equation}
\end{shrinkeq}
Then, we have
\begin{equation*}
\begin{aligned}
  \widetilde {{C_{{l_1}}}}(\rho ) & = \sum\limits_{k = 1}^n {{q_k}{C_{{l_1}}}(|{\phi _k}\rangle )}  = \sum\limits_{k = 1}^n {{q_k}\sum\limits_{i \ne j} {{\phi _{ki}}{\phi _{kj}}} }  \\
& \ge \sum\limits_{k = 1}^n {{q_k}\frac{{\sum\limits_{i \ne j} {{\phi _{ki}}{\phi _{kj}}} }}{{1 + \sum\limits_{i \ne j} {{\phi _{ki}}{\phi _{kj}}} }}}  \ge \frac{{\sum\limits_{k = 1}^n {q_k^2
\sum\limits_{i \ne j} {{\phi _{ki}}{\phi _{kj}}} } }}{{\sum\limits_{k = 1}^n {{q_k}(1 + \sum\limits_{i \ne j} {{\phi _{ki}}{\phi _{kj}}} )} }}  \\
&  \ge \frac{{\sum\limits_{k = 1}^n {q_k^2\sum\limits_{i \ne j} {{\phi _{ki}}{\phi _{kj}}} } }}{{\sum\limits_{k = 1}^n {{p_k}{C_{{l_1}}}(|{\psi _k}\rangle )}  + 1}}
= \frac{{\sum\limits_{k = 1}^n {\sum\limits_{i \ne j} {{{\left( {\sqrt {q_k^{}{\phi _{ki}}{\phi _{kj}}} } \right)}^2}} } }}{{\sum\limits_{k = 1}^n {{p_k}{C_{{l_1}}}(|{\psi _k}\rangle )}  + 1}}  \\
&  \ge \frac{{\sum\limits_{i \ne j} {{{\left( {\sum\limits_{k = 1}^n {\sqrt {q_k^{}{\phi _{ki}}{\phi _{kj}}} } } \right)}^2}} }}{B}  \ge \frac{{\sum\limits_{i \ne j} {{{\left| {{\rho _{ij}}}
\right|}^2}} }}{B}\ge \frac{{{{\left| {{\rho _{st}}} \right|}^2}}}{B}.
\end{aligned}
\end{equation*}
\hfill \rule{1ex}{1ex}

We can also prove that the upper bound provided by theorem
\ref{dl4} is better than that provided by Lemma
\ref{yl8}.
\begin{theorem}   
If a mixed state $\rho$ can be expressed as a convex combination of $n$ pure states, $|{\psi _1}\rangle ,|{\psi _2}\rangle , \cdots ,|{\psi _n}\rangle$, that is,
\begin{equation*}
  \rho  = \sum\nolimits_{i = 1}^n {{p_i}|{\psi _i}\rangle \langle {\psi _i}|},
\end{equation*}
then it holds that
\begin{equation*}
\begin{aligned}
\frac{1}{n}\sum\limits_{k = 1}^n {\left[ {\widetilde G_k^{(n - 1)} + {p_k}\widetilde {{C_{{l_1}}}}(|{\psi _k}\rangle )} \right]}  \le \sum\limits_{k = 1}^n {{p_k}\widetilde {{C_{{l_1}}}}(|{\psi _k}\rangle )}.
\end{aligned}
\end{equation*}
\end{theorem}

{\bf{Proof:}}
Assume that a mixed state $\rho$ can be expressed as a convex combination of $n$ pure states $|{\psi _1}\rangle ,|{\psi _2}\rangle , \cdots ,|{\psi _n}\rangle$, that is,
\begin{equation*}
  \rho  = \sum\nolimits_{i = 1}^n {{p_i}|{\psi _i}\rangle \langle {\psi _i}|}.
\end{equation*}
Thus, we have
\begin{equation*}
\begin{aligned}
&  \quad \ \frac{1}{n}\sum\limits_{k = 1}^n {\left[ {\widetilde G_k^{(n - 1)} + {p_k}\widetilde {{C_{{l_1}}}}(|{\psi _k}\rangle )} \right]} \\
&  = \frac{1}{n}\sum\limits_{k = 1}^n {\widetilde G_k^{(n - 1)}}  + \frac{1}{n}\sum\limits_{k = 1}^n {{p_k}\widetilde {{C_{{l_1}}}}(|{\psi _k}\rangle )}  \\
&  = \frac{1}{n}\sum\limits_{k = 1}^n {(1 - {p_k})\widetilde {{C_{{l_1}}}}(\sum\limits_{i \ne k} {{\textstyle{\frac{{p_i}}{1 - {p_k}}}}|{\psi _i}\rangle } )}
 + \frac{1}{n}\sum\limits_{k = 1}^n {{p_k}\widetilde {{C_{{l_1}}}}(|{\psi _k}\rangle )} \\
& \le \frac{1}{n}\sum\limits_{k = 1}^n {(1 - {p_k})\sum\limits_{i
\ne k} {{\textstyle{\frac {{p_i}}  {1 - {p_k}}}}\widetilde
{{C_{{l_1}}}}(|{\psi _i}
\rangle )} } +  \frac{1}{n}\sum\limits_{k = 1}^n {{p_k}\widetilde {{C_{{l_1}}}}(|{\psi _k}\rangle )}  \\
&  = \frac{{(n - 1)}}{n}\sum\limits_{k = 1}^n {{p_k}\widetilde {{C_{{l_1}}}}(|{\psi _k}\rangle )}  + \frac{1}{n}\sum\limits_{k = 1}^n {{p_k}\widetilde {{C_{{l_1}}}}(|{\psi _k}\rangle )} \\
& = \sum\limits_{k = 1}^n {{p_k}\widetilde {{C_{{l_1}}}}(|{\psi _k}\rangle )}.
\end{aligned}
\end{equation*}
\hfill \rule{1ex}{1ex}

\section{Different upper bounds of coherence}
In this section, we detail the evaluation of the coherence of a
quantum state using its properties. Ref.\cite{16}, using the
dimension of a quantum state, recorded the following conclusion:
\begin{lemma} \label{l8}   
For any $d$-dimensional quantum state $\rho$, it holds that
${C_{{l_1}}}(\rho ) \le d - 1$. Moreover, ${C_{{l_1}}}(\rho )$
reaches the upper bound when $\rho$ is in the maximally mixed state:
\begin{equation*}
    |\psi \rangle  = \frac{1}{{\sqrt d }}\sum\nolimits_{i = 1}^d {|i\rangle }.
\end{equation*}
\end{lemma}

However, in some cases, the evaluation provided by Lemma \ref{l8} is
highly inaccurate. For example,
\begin{equation*}
  \rho  = diag\{ {\lambda _1},{\lambda _2}, \cdots ,{\lambda _d}\}.
\end{equation*}

With additional conditions, Lemma \ref{l8} can be improved. To this end,
we introduce the following lemma \cite{20}.
\begin{lemma}    \label{l10}
Consider that $A = [{a_{ij}}]$ is positive semi-definite and that for some $k \in \{ 1, \cdots ,n\}$, it holds that ${a_{kk}} = 0$. Then, for each $i \in\{ 1, \cdots ,n\}$,
we have ${a_{ik}} = {a_{ki}} = 0$.
\end{lemma}

\begin{theorem} 
Let $\rho$ be a $d$-dimensional quantum state. If there exists $r$
null elements in the main diagonal of $\rho$, then
\begin{equation}
\begin{aligned}
&{C_{{l_1}}}(\rho ) \le d - r - 1.
\end{aligned}
\end{equation}
\end{theorem}

{\bf{Proof:}}
Assume that $\rho$ is a $d$-dimensional quantum state satisfying
\begin{equation*}
  {\rho _{{k_i},{k_i}}} \neq 0,i = 1,2, \cdots ,d - r({k_1} < {k_2} <  \cdots  < k_{d - r}).
\end{equation*}
Let the matrix of $\rho$ under reference basis $\left\{ {|i\rangle }
\right\}_{i = 1}^d$ be
\begin{equation*}
\rho  = \left( {\begin{array}{*{20}{c}}
{{\rho _{11}}}&{{\rho _{12}}}& \cdots &{{\rho _{1d}}}\\
{{\rho _{21}}}&{{\rho _{22}}}& \cdots &{{\rho _{2d}}}\\
 \vdots & \vdots &{}& \vdots \\
{{\rho _{d1}}}&{{\rho _{d2}}}& \cdots &{{\rho _{dd}}},
\end{array}} \right).
\end{equation*}

Define
\begin{equation*}
\rho ' = \left( {\begin{array}{*{20}{c}}
{{\rho _{{k_1},{k_1}}}}& * & \cdots & * \\
 * &{{\rho _{{k_2},{k_2}}}}&{}& * \\
 \vdots &{}& \ddots & \vdots \\
 * & * & \cdots &{{\rho _{{k_{d - r}},{k_{d - r}}}}}
\end{array}} \right).
\end{equation*}
where $\rho'$ is obtained via the permutation of each element of
$\rho$ whose row index or column index is $k_i, i=1,2,\cdots,d-r$
in the same order. Furthermore, all the
principal minors of $\rho'$ are the principal minors of $\rho$.
Therefore, all the principal minors of $\rho'$ are non-negative, given that $\rho$ is positive semi-definite.

In addition,
\begin{equation*}
  tr(\rho') = tr(\rho) = 1.
\end{equation*}
Hence, $\rho'$ is a density matrix, the dimension of which is $d - r$.

According to Lemma \ref{l8} and Lemma \ref{l10}, we can directly
obtain
\begin{equation*}
  {C_{{l_1}}}(\rho ) = {C_{{l_1}}}({\rho'}) \leq d - r - 1.
\end{equation*}
\hfill \rule{1ex}{1ex}

Furthermore, we provide another two distinct upper bounds of the $l_{1}$-norm.
\begin{theorem}  
Assume that $\rho$ is a $d$-dimensional quantum state. For convenience, the matrix of $\rho$ under the reference basis $\left\{ {|i\rangle } \right\}_{i = 1}^d$ is still denoted by $\rho$.
 All integers $k (2\leq k \leq d)$ are assembled such that ${{\rho _{k,1}},{\rho _{k + 1,2}}, \cdots {\rho _{d,d - k + 1}}}$ are not complete zeros in a set ${S_\rho }$.
Defining the number of elements in ${S_\rho }$ as $t$, we have
\begin{equation*}
{C_{{l_1}}}(\rho ) \le 2t.
\end{equation*}
\end{theorem}

{\bf{Proof:}}
Assume that $\rho$ is a $d$-dimensional quantum state. Let
\begin{equation*}
  {S_\rho } =\left\{ {r_{1},r_{2},\cdots,r_{t}} \right\}.
\end{equation*}

All the principal minors of $\rho$ are non-negative because $\rho$ is positive semi-definite, that is,
\begin{equation*}
\begin{aligned}
&\rho \left( {\begin{array}{*{20}{c}}
i&j\\
i&j
\end{array}} \right) = \left| {\begin{array}{*{20}{c}}
{{\rho _{ii}}}&{{\rho _{ij}}}\\
{{\rho _{ji}}}&{{\rho _{jj}}}
\end{array}} \right| = {\rho _{ii}}{\rho _{jj}} - {\left| {{\rho _{ij}}} \right|^2} \ge
0.
\end{aligned}
\end{equation*}
Furthermore, we have
\begin{equation*}
\begin{aligned}
&\left| {{\rho _{ij}}} \right| \le \sqrt {{\rho _{ii}}{\rho _{jj}}}
\le \frac{1}{2}({\rho _{ii}} + {\rho _{jj}}).
\end{aligned}
\end{equation*}

It should be noted that $\rho$ is Hermite; hence, it follows that
\begin{equation*}
\begin{aligned}
{C_{{l_1}}}(\rho )& = \sum\limits_{i \ne j} {\left| {{\rho _{ij}}}
\right|}  = 2\sum\limits_{i = 1}^t
 {\sum\limits_{j = 1}^{d - {r_i} + 1} {\left| {{\rho _{{r_i} + j - 1,j}}} \right|} }  \\
&\leq 2\sum\limits_{i = 1}^t {\sum\limits_{j = 1}^{d - {r_i} + 1} {\frac{{\left( {{\rho _{{r_i} + j - 1,{r_i} + j - 1}} + {\rho _{jj}}} \right)}}{2}} }\\
& \le 2{\sum\limits_{j = 1}^d {{\textstyle{\frac{{\rho _{jj}}}{
2}}}} }  + 2\sum\limits_{i = 1}^t {\sum\limits_{j = 1}^{d - {r_i} +
1} {\frac{{{\rho _{{r_i} + j - 1,{r_i} + j - 1}}}}{2}} }\\ & = t +
\sum\limits_{i = 1}^t {\sum\limits_{j = 1}^{d - {r_i} + 1} {{\rho
_{{r_i} + j - 1,{r_i} + j - 1}}} }.
\end{aligned}
\end{equation*}

We declare that, among the following $t(d + 1) - ({r_1} + {r_2} +  \cdots  + {r_t})$ numbers
\begin{equation*}
\begin{aligned}
& \rho _{{r_1},{r_1}},\rho _{{r_1} + 1,{r_1} + 1}, \cdots ,\rho _{d,d}, \cdots , \\
&\quad \rho _{{r_t},{r_t}},\rho _{{r_t} + 1,{r_t} + 1}, \cdots ,\rho _{d,d},
\end{aligned}
\end{equation*}
the same one appears $t$ times at most; thus,
\begin{equation*}
\begin{aligned}
{C_{{l_1}}}(\rho ) & \le t + \sum\limits_{i = 1}^t {\sum\limits_{j = 1}^{d - {r_i} + 1} {{\rho _{{r_i} + j - 1,{r_i} + j - 1}}} } \\
&\le t + t({\rho _{11}} +  \cdots  + {\rho _{dd}}) = 2t.
\end{aligned}
\end{equation*}
\hfill \rule{1ex}{1ex}

\begin{theorem}   
Let $\rho$ be a $d$-dimensional $(d\geq2)$ quantum state. If there
exists ${\rho _{ij}}$ such that $|{\rho _{ij}}| = 0$, then we have
\begin{equation}
\begin{aligned}
&{C_{{l_1}}}(\rho ) \le \sqrt {\left[ {d(d - 1) - 1} \right]\mu }
\le d - 1,
\end{aligned}
\end{equation}
where $\mu  = \sum\limits_{k = 1}^d {\lambda _k^2}  - \sum\limits_{k = 1}^d {\rho _{_{kk}}^2}$ and ${\lambda _k}\left( {k = 1,2, \cdots ,d} \right)$ are the eigenvalues of $\rho$.
\end{theorem}

{\bf{Proof:}} Let $\rho$ be a $d$-dimensional quantum state.
According to the spectral decomposition theorem, there exists a
unitary matrix $U$ such that
\begin{equation*}
\begin{aligned}
&\rho  = U\left( {\begin{array}{*{20}{c}}
{{\lambda _1}}&{}&{}&{}\\
{}&{{\lambda _2}}&{}&{}\\
{}&{}& \ddots &{}\\
{}&{}&{}&{{\lambda _d}},
\end{array}} \right){U^ + }.
\end{aligned}
\end{equation*}
Then, $\rho^{2}$ can be expressed as
\begin{equation*}
\begin{aligned}
{\rho ^2} & = U\left( {\begin{array}{*{20}{c}}
{{\lambda _1}}&{}&{}\\
{}& \ddots &{}\\
{}&{}&{{\lambda _d}}
\end{array}} \right){U^ + }U\left( {\begin{array}{*{20}{c}}
{{\lambda _1}}&{}&{}\\
{}& \ddots &{}\\
{}&{}&{{\lambda _d}}
\end{array}} \right){U^ + } \\
& = U\left( {\begin{array}{*{20}{c}}
{\lambda _{\rm{1}}^{\rm{2}}}&{}&{}\\
{}& \ddots &{}\\
{}&{}&{\lambda _d^2}
\end{array}} \right){U^ + }.
\end{aligned}
\end{equation*}
Following the expression of $\rho^{2}$ established above, we obtain
\begin{equation*}
\begin{aligned}
&tr({\rho ^2}) = \sum\limits_{i = 1}^d {\sum\limits_{j = 1}^d {|{\rho _{ij}}{|^2}} }  = \sum\limits_{k = 1}^d {\lambda _k^2}.
\end{aligned}
\end{equation*}
Furthermore, we have
\begin{equation*}
\begin{aligned}
&\sum\limits_{i \ne j} {|{\rho _{ij}}{|^2}}  = \sum\limits_{k = 1}^d {\lambda _k^2}  - \sum\limits_{k = 1}^d {\rho _{_{kk}}^2}.
\end{aligned}
\end{equation*}

Now, we introduce the modified Cauchy inequality \cite{21}. Assume that
$V$ is a product space; thus, for any $x,y,z \in V$, where $\|z\| =
1$, it holds that
\begin{equation*}
\begin{aligned}
&{\left| {(x,y)} \right|^2} \le {\left\| x \right\|^2}{\left\| y
\right\|^2} - G(x,y,z),
\end{aligned}
\end{equation*}
where
\begin{equation*}
  G(x,y,z) = {\left( {\left\| x \right\|(y,z) - \left\| y \right\|(x,z)} \right)^2}.
\end{equation*}
The equality holds only if $x,y,z$ are linearly dependent.

Now, we attempt to determine an upper bound of ${C_{{l_1}}}(\rho )$ using this useful tool. Let $V = {\mathbb{R}^{n}}(n \ge 2)$ and the standard inner product be taken as the fixed product. Subsequently, $x,y,z$ is selected in the following manner:
\begin{equation*}
\begin{aligned}
&x = ({x_1},{x_2}, \cdots ,{x_n}),{\rm{ }}y = (\frac{1}{{\sqrt n }},\frac{1}{{\sqrt n }}, \cdots ,\frac{1}{{\sqrt n }}), \\
& \qquad \qquad  {\rm{ }}z = (0,0, \cdots ,1,0, \cdots ,0).
\end{aligned}
\end{equation*}
where the $i$th element of the $n$-tuple $z$ is 1. Owing to the definition of the standard inner product, we have  
\begin{equation*}
\begin{aligned}
&(x,y) = \frac{1}{{\sqrt n }}\sum\limits_{i = 1}^n {{x_i}} ,\left\|
x \right\| =
\sqrt {\sum\limits_{i = 1}^n {x_i^2} } ,\left\| y \right\| = 1, \\
&\qquad \qquad (y,z) = \frac{1}{{\sqrt n }},(x,z) = {x_i}.
\end{aligned}
\end{equation*}
Using the modified Cauchy inequality, we have
\begin{equation*}
\begin{aligned}
&\frac{1}{n}{\left( {\sum\limits_{i = 1}^n {{x_i}} } \right)^2} \le \sum\limits_{i = 1}^n {x_i^2}.
{\left( {\sqrt {\sum\limits_{i = 1}^n {x_i^2} } \frac{1}{{\sqrt n }} - {x_i}} \right)^2}.
\end{aligned}
\end{equation*}

For the convenience of narration, we introduce the following notations:
\begin{equation*}
\begin{aligned}
&\sum\limits_{i = 1}^n {{x_i}} : = u,\sum\limits_{i = 1}^n {x_i^2} : = \lambda.
\end{aligned}
\end{equation*}
Then,
\begin{equation*}
\begin{aligned}
&\frac{1}{n}{u^2} \le \lambda  - {\left( {\sqrt \lambda  \frac{1}{{\sqrt n }} - {x_i}} \right)^2}.
\end{aligned}
\end{equation*}
In other words,
\begin{equation}
\begin{aligned}
&\frac{1}{n}{u^2} - \lambda  + \frac{1}{n}\lambda  \le \frac{2}{{\sqrt n }}\sqrt \lambda  {x_i} - x_i^2.  \label{11}
\end{aligned}
\end{equation}
Note that Eq.(\ref{11}) holds for each $i$; hence,
\begin{equation*}
  \frac{1}{n}{u^2} - \lambda  + \frac{1}{n}\lambda  \le \min \left\{ {\left. {\sqrt {\frac{{{\rm{4}}\lambda }}{n}} {x_i} - x_i^2} \right|i = 1, \ ldots ,n} \right\}.
\end{equation*}
If $j \in \{ 1,2, \cdots ,n\}$, then ${x_j} = 0$. Therefore,
\begin{equation*}
\begin{aligned}
&\frac{1}{n}{u^2} - \lambda  + \frac{1}{n}\lambda  \le 0.
\end{aligned}
\end{equation*}
In other words,
\begin{equation*}
\begin{aligned}
    u\leq\sqrt {(n - 1)\lambda }.
\end{aligned}
\end{equation*}
In summary, it holds that
\begin{equation*}
\begin{aligned}
&{C_{{l_1}}}(\rho )=\sum\limits_{i \ne j} {|{\rho _{ij}}|}  \le
\sqrt {\left[ {d(d - 1) - 1} \right]\mu },
\end{aligned}
\end{equation*}
where $\mu  = \sum\limits_{k = 1}^d {\lambda _k^2}  - \sum\limits_{k,
= 1}^d {\rho _{_{kk}}^2}.$

Next, we focus on
\begin{equation*}
\begin{aligned}
&\sqrt {\left[ {d(d - 1) - 1} \right]\mu }  \le d - 1.
\end{aligned}
\end{equation*}
According to the properties of density matrix, we have
\begin{equation*}
\begin{aligned}
&\ \  \sum\limits_{k = 1}^d {\lambda _k^2}  \le {\left( {\sum\limits_{k = 1}^d {\lambda _k^{}} } \right)^2} = 1,  \\
&\sum\limits_{k = 1}^d {\rho _{_{kk}}^2}  \ge \frac{1}{d}{\left( {\sum\limits_{k = 1}^d {\rho _{_{kk}}^{}} } \right)^{\rm{2}}} = \frac{1}{d}.
\end{aligned}
\end{equation*}
Thus, $\mu$ can be expressed as
\begin{equation}
\begin{aligned}
&\mu  \le 1 - \frac{1}{d} = \frac{{d - 1}}{d}. \label{12}
\end{aligned}
\end{equation}
From Eq.(\ref{12}), we can directly obtain
\begin{equation*}
\begin{aligned}
\sqrt {\left[ {d(d - 1) - 1} \right]\mu }  &\le \sqrt {\left[ {d(d - 1) - 1} \right]\frac{{d - 1}}{d}} \\
& = \sqrt {{{\left( {d - 1} \right)}^2} - \frac{{d - 1}}{d}}  \le d - 1.
\end{aligned}
\end{equation*}
\hfill \rule{1ex}{1ex}

\section{Conclusion and discussion}

We studied the properties of quantum coherence based on the
$l_{1}$-norm and convex-roof $l_{1}$-norm, thus realizing a better generalized triangular
inequality. Furthermore, we proved that, for a certain type of three-dimensional
quantum state $\rho$, it holds that ${C_{{l_1}}}\left( \rho \right)
= \widetilde {{C_{{l_1}}}}\left( \rho  \right)$. In addition, we offer a few upper bounds for the $l_{1}$-norm in different
forms according to the properties of a quantum state.

For future research, we propose the following conjecture: Let
$\rho$ be a three-dimensional quantum state with $rank(\rho)\neq2$; then,
\begin{equation*}
  {C_{{l_1}}}(\rho ) = \widetilde {{C_{{l_1}}}}(\rho ).
\end{equation*}
The difficulty lies in calculating the convex-roof
$l_{1}$-norm of more general quantum states.

We presented certain previously unreported facts,
which we believe will be helpful in the quantitative estimation of quantum
coherence and other quantum resources.

\bigskip
Acknowledgments: This work is supported by Hainan Provincial Natural
Science Foundation of China under Grant Nos.121RC539 and 121MS030;
and by the National Natural Science Foundation of China under Grant
No.11861031. This project is also supported by the Innovation
Platform for Academinicians of Hainan Province


\bigskip

\begin{thebibliography}{99}
\bibitem{1}  A. Anurag, D. V. Krishna and J.Rahul, Quantum communication using coherent rejection sampling, Phys. Rev. Lett. \textbf{119}, 120506 (2017).
\bibitem{2}  J. Mompart, K. Eckert, W. Ertmer, G. Birkl and L. M, Quantum computing with spatially delocalized qubits, Phys. Rev. Lett. \textbf{90}, 147901 (2003).
\bibitem{3}  M. Gonzalo, P. Francesco and Z. Roberta, Optimal work extraction and thermodynamics of quantum measurements and correlations, Phys. Rev. Lett. \textbf{121}, 120602 (2018).
\bibitem{4}  K. Matsuno, Forming and maintaining a heat engine for quantum biology, BioSystems {\bf 85}, 23 (2006).
\bibitem{5}  V. Anisimov and J. P. Stewart, Introduction to Quantum Biology[M]. CRC Press Inc Bosa Roca, (2015).
\bibitem{6}  S. Naoto and S. Takahiro, Quantum thermodynamics of correlated-catalytic state conversion at small scale, Phys. Rev. Lett. \textbf{126}, 150502 (2021).
\bibitem{7}  V. Narasimhachar and G.Gour, Low-temperature thermodynamics with quantum coherence, Nature Communications {\bf 6},7689(2015).
\bibitem{8}  G. Benenti, G. Casati, S. Montangero and D. L. Shepelyansky, Efficient quantum computing of complex dynamics, Phys. Rev. Lett. \textbf{87}, 227901 (2001).
\bibitem{9} A. Sarkar, Z. Ai-Ars and K. Bertels, Estimating algorithmic information using quantum computing for genomics applications, Applied Sciences \textbf{11}, 2696 (2021).
\bibitem{10} T. Baumgratz, M. Cramer and M. B. Plenio, Quantifying coherence, Phys. Rev. Lett. \textbf{113}, 140401 (2014).
\bibitem{11} N. Carmine, T. R. Bromley, C. Marco, P. Marco, J. Nathaniel and A. Gerardo, Robustness of coherence: an operational and observable measure of quantum coherence, Phys. Rev. Lett. \textbf{116}, 150502 (2016).
\bibitem{12} K. Bu, S. Uttam, S. M. Fei, P. A. Kumar and J. D. Wu, Maximum relative entropy of coherence: an operational coherence measure, Phy. Rev. Lett. \textbf{119}, 150405 (2017).
\bibitem{13} S. Alexander, S. Uttam, D. H. Shekhar, B. M. Nath and A. Gerardo, Measuring quantum coherence with entanglement, Phys. Rev. Lett. \textbf{115}, 020403 (2015).
\bibitem{mjmt} M. J. Zhao, T. Ma and R. Pereira, Average quantum coherence of pure-state decomposition, Phys. Rev. A {\bf 103}, 042428 (2021).
\bibitem{sdru} S. Designolle, R. Uola, K. Luoma, and N. Brunner, Set coherence: basis-independent quantification of quantum coherence, Phys. Rev. Lett. {\bf 126}, 220404 (2021).
\bibitem{fcle} X. Chen, et al. Generalized multipath wave-particle duality in a delayed-choice experiment, Nature Commun. {\bf 12}, 2712 (2021).
\bibitem{14} Y. Dai, W. L. You, Y. L. Dong and C. J. Zhang, Triangle inequalities in coherence measures and entanglement concurrence, Phys. Rev. A \textbf{96}, 062308 (2017).
\bibitem{15} Z. X. Jin, X. Li-Jost and S. M. Fei, Triangle-like inequalities related to coherence and entanglement negativity, Quantum Inf Process, {\bf 18}, 5 (2019).
\bibitem{16} Z. Jiang, T. Zhang, X. Huang and S. M. Fei, Trade-off relations of $l_1$-norm coherence for multipartite systems, Quantum Inf Process, \textbf{19}, 92 (2020).
\bibitem{17} X. F. Qi, T. Gao and F. L. Yan, Measuring coherence with entanglement concurrence, J. Phys. A: Math. Theor. \textbf{50}, 285301 (2017).
\bibitem{18} X. Yuan, H. Zhou, Z. Cao and X. Ma, Intrinsic randomness as a measure of quantum coherence, Phys. Rev. A \textbf{92}, 022124
(2015).
\bibitem{19} M. J. Zhao, T. Ma, Z. Wang, S. M. Fei and R. Pereira, Coherence concurrence for X states, Quantum Inf Process,  {\bf 19}, 3 (2020).
\bibitem{20} R. A. Horn and C. R. Johnson, Matrix Analysis[M]. Cambridge: Cambridge University Press, (2013).
\bibitem{21} M. Gao, On Hilbert's integral inequality, Mathematica Applicata, {\bf 11(3)}, 32-35 (1998).
\end{thebibliography}
\end{document}